\documentclass[aps,pra,showpacs,twocolumn,preprint, 10pt]{revtex4}

\usepackage{graphicx}

\usepackage{color}

\begin{document}

\title{Nonclassical characteristic functions for highly sensitive measurements}

\author{Th.~Richter and W.~Vogel}

\address{Arbeitsgruppe Quantenoptik, Fachbereich  Physik,
  Universit\"at Rostock, D-18051 Rostock, Germany}

\date{\today}

\begin{abstract}
  Characteristic functions are shown to be useful for highly sensitive
  measurements.  Redistributions of motional Fock states of a trapped atom can
  be directly monitored via the most fragile nonclassical part of the
  characteristic function. The method can also be used for decoherence
  measurements in optical quantum-information systems.
\end{abstract}

\pacs{03.65.Bz, 42.50.Ar, 42.50.Dv}

\maketitle

\section{Introduction}

The experimental demonstrations of photon antibunching~\cite{kimble},
sub-Poissonian photon statistics~\cite{short} and quadrature
squeezing~\cite{slusher} also led to an increasing interest in practical
applications of nonclassical states.  An early example is the proposal to use
squeezed light for enhancing the sensitivity of interferometric
gravitational-waves detection~\cite{caves}. Experiments have demonstrated the
usefulness of squeezed light for improving interferometric
measurements~\cite{xiao,grangier} and spectroscopy~\cite{polzik}.

Two decades after the first experimental demonstrations of the
potential usefulness of nonclassical states the latter still play a minor role
in practical measurements. There may be several reasons for this fact.
First, the experimental effort for generating the needed nonclassical
states is rather high. Second, some applications, e.g.
the use of squeezed light for optimizing the laser power in gravitational-wave
detection, can be replaced with developments of laser sources.
Third, nonclassical states are usually highly fragile against losses which may
substantially limit their advantages in some applications.

The use of nonclassical states is frequently considered in the context of the
reduction of the quantum noise in a certain observable below an ultimate
classical noise limit.  Examples are the use of sub-Poissonian and
squeezed light fields for reducing the noise in direct and homodyne
photodetection, respectively. This requires to link the
measurement principle with the observable whose quantum noise is reduced.
Below we will reconsider the application of 
nonclassical states from a much broader point of view.
When speaking about nonclassical states in the following, we will only
consider quantum states of a single-mode harmonic oscillator whose
Glauber-Sudarshan $P$-function is not a probability density~\cite{titulaer}.

The nonclassicality of quantum states can be completely characterized in terms
of measurable characteristic functions of phase-dependent quadratures,
\begin{equation}
  \label{eq:xop}
  \hat{x}_\varphi = \hat{a} \, e^{i\varphi} + \hat{a}^\dagger
  e^{-i\varphi} ,
\end{equation}
$\hat{a}$ ($\hat{a}^\dagger$) being the bosonic annihilation (creation)
operator and $\varphi$ is the phase parameter.
To be more specific, a hierarchy of necessary and sufficient
conditions has been derived that completely characterizes the nonclassicality
of a given quantum state in terms of the quadrature characteristic function
$G(k,\varphi)$~\cite{ri-vo}.  A broad class of nonclassical states can be well
characterized by the rather simple condition of first-order nonclassicality,
\begin{equation}
\label{eq:first-order}
|G(k,\varphi)| \ge G_{\rm gr}(k) ,
\end{equation}
stating that the absolute value of the characteristic function exceeds, for some arguments,
the corresponding value of the ground (or vacuum) state~\cite{vogel}.

The signatures of first-order nonclassicality are more general than the
quantum-noise reduction of a chosen observable below some classical limit.
The condition also includes features like quantum interference~\cite{vowe} and
sub-Planck structures in phase space~\cite{zurek}.  Note that the
nonclassical effects of first and second order have been experimentally demonstrated for
radiation fields~\cite{lvovsky,zavatta}. The needed characteristic
functions can also be observed for the quantized center-of-mass motion of a
trapped ion~\cite{wavo,monroe}.

In this paper we propose a new measurement principle, where the quadrature
characteristic function serves as a highly sensitive probe.  It makes use of
the fact that the nonclassical signatures of the quadrature characteristic
function are more fragile with respect to dissipation than other nonclassical
effects, such as sub-Poissonian statistics and squeezing.  This provides
a new tool for the highly sensitive diagnostics of decoherence effects, which
is of great interest for quantum information processing. 

The paper is organized as follows. In Sec.~II we consider the detection of the
characteristic function of the quadrature distribution for the motion of
trapped ions and for propagating radiation fields. The decoherence is caused
in both cases, for example, by a thermal reservoir. Section III is devoted to
the use of the most pronounced nonclassical features of the characteristic
functions for highly sensitive measurements. A brief summary is given in
Sec.~IV.

\section{Decoherence in Terms of Characteristic Functions}

Let us consider a nonclassical state $\hat{\rho}(0)$ of a bosonic mode 
prepared at the initial time $t=0$. Its further evolution is caused by the
dissipation to be analyzed, leading to the state $\hat{\rho}(t)$. Eventually,
the quadrature characteristic function $G(k,t,\varphi)$ is measured.

In the case of a trapped ion the measurement of $G(k,t,\varphi)$ is performed
as follows. An electronic transition is driven simultaneously
on the red and the blue motional sidebands in the resolved sideband
regime, which is described by the interaction Hamiltonian~\cite{wavo}
\begin{equation}
  \label{eq:ham}
  \hat{H}_{\rm int} = \hbar \left( \Omega \, \hat{A}_{12} +
    \Omega^\ast \hat{A}_{21} \right) \hat{x}_\varphi ,
\end{equation}
where $\hat{A}_{ij}=|i\rangle \langle j |$ ($i,j=1,2$) is the electronic flip
operator and $\Omega$ is the effective Rabi frequency. Most importantly, it is
proportional to the quadrature operator $\hat{x}_\varphi$ of
the center-of-mass motion, the phase $\varphi$ being controlled by the phase
difference of the driving lasers.
The total state of the ion is $\hat{\varrho}(t) = \hat{\rho}(t) \otimes
\hat{\sigma}(t)$, with a properly prepared electronic state
$\hat{\sigma}(t)$.  At time $t$ the interaction~(\ref{eq:ham}) is switched on
for the interaction time $\tau$.  The observation of the occupation
${\sigma}_{11}(t+\tau,\varphi)$ of the electronic ground state directly yields
the characteristic function $G(k,t,\varphi)$ of the quadrature
distribution~\cite{wavo}:
\begin{eqnarray} \label{eq:ini}
G(k,t,\varphi) = 
2 \left [\sigma^{(\rm inc)}_{11}(t+\tau,\varphi)- {\textstyle\frac{1}{2}} \right]\\\nonumber 
 +2i \left [\sigma^{(\rm coh)}_{11}(t+\tau,\varphi)- {\textstyle\frac{1}{2}} \right]\,,
\end{eqnarray}
where the interaction~(\ref{eq:ham}) leads to the scaling
$k =2 |\Omega| \tau$.  The incoherent and coherent
occupations, $\sigma^{(\rm inc)}_{11}$ and $\sigma^{(\rm coh)}_{11}$, are
measured with the electronic preparations ${\sigma}_{11}(t) \!=\! 1$
and $\sigma_{11}(t) \!=\!  |\sigma_{12}(t)| \!=\!  \frac{1}{2}$, respectively.
The electronic-state occupations in Eq.~(\ref{eq:ini}) are detected with
almost perfect efficiency, by testing (at time $t+\tau$) a transition from the
state $|1\rangle$ to
an auxiliary state for the appearance of fluorescence~\cite{monroe}.

For a radiation field the characteristic function can be sampled in 
balanced homodyne detection~\cite{lvovsky,zavatta},
\begin{equation}
\label{CF-sampl}
G(k,t,\varphi) \approx \frac{1}{N_\varphi} \sum_{j=1}^{N_\varphi} e^{ik x_{\varphi,j}(t)}.
\end{equation} 
Here $x_{\varphi,j}(t)$, $j=1,\dots, N_\varphi$, is the set of data recorded
in balanced homodyning for each setting of $\varphi$. The quadratures describe
now the radiation mode at the time $t$. Thus they carry the information on the
dissipative interaction contained in the radiation-state $\hat{\rho}(t)$.
 
To illustrate the idea, we deal with a simple model of decoherence caused by a
thermal bath of mean occupation number $\bar{n}$. The density operator
$\hat{\rho}$ in the interaction picture
obeys the master equation
\begin{eqnarray}\label{master-eq}
\frac{d}{dt}\hat{\rho}& = &
\gamma\,(\bar{n}+1)\,[2\hat{a}\hat{\rho}\hat{a}^{\dagger} -
\hat{a}^{\dagger}\hat{a}\hat{\rho} - \hat{\rho}\hat{a}^{\dagger}\hat{a}]
\nonumber \\
& & + \gamma\,\bar{n}\,[2\hat{a}^{\dagger}\hat{\rho}\hat{a} -
\hat{a}\hat{a}^{\dagger}\hat{\rho} -
\hat{\rho}\hat{a}\hat{a}^{\dagger}]\,,              
\end{eqnarray}
with $\gamma$ being the damping rate.  
The resulting equation for the Wigner characteristic function, $\chi(\xi,t)
\equiv Tr\left\{\hat{\varrho} \exp(\xi
  \hat{a}^{\dagger}-\xi^{\ast}\hat{a})\right\}$, has the
solution~\cite{marian}
\begin{eqnarray}\label{char-eq}
\chi(\xi,t) & = 
& \exp\left\{-(\bar{n}+1/2)\,|\xi|^2\,[1-\exp(-2\gamma
  t)]\right\} \nonumber \\
& &  \times\chi(\xi\exp(-\gamma t),0)  \, ,
\end{eqnarray}
where $\chi(\xi,0)$ is the Wigner characteristic function of the initial
quantum state. From this result,
the observable quadrature characteristic function, 
\begin{equation}\label{GFTW}
G(k,t,\varphi)= \chi(ike^{-i\varphi},t),
\end{equation}
is easily derived.

\section{Highly sensitive Detection}

\subsection{Decoherence of a Trapped Ion}

Let us first consider a trapped atom which is initially in the
number state $|m\rangle$, which can be realized in
experiments~\cite{meekhof}. 
The motional-state redistributions caused by the reservoir 
lead to strong modifications of the nonclassical
signatures of the characteristic function, which represents our highly
sensitive probe. Note that the observed decoherence of a
Raman-driven trapped ion~\cite{meekhof} is not
completely understood yet. Although dephasing mechanisms could be
identified~\cite{difidio}, a deeper insight in the role of motional states is
still required.

The Wigner characteristic function of the number state $|m\rangle$ 
is given by
\begin{equation}
\label{CF-n}
\chi_m(\xi)  = L_m(|\xi|^2) \,\exp{(-|\xi|^2/2)}, 
\end{equation}
where $L_m(x)$ is a Laguerre polynomial of order $m$.  Since the state
$|m\rangle$ is phase independent, we may write $G_{\rm m}(k,\varphi)= G_{\rm
  m}(k)$.  In Fig.~1 we show the quadrature characteristic functions, $G_{\rm
  m}(k)\equiv G_{\rm m}(k,t=0)$, as functions of $k$, for $m=9,10,11$.
Clearly, the first-order nonclassicality condition~(\ref{eq:first-order}) is
fulfilled for all the shown number states.  For our examples, the first-order
nonclassical effect is most pronounced at the outermost local extremum, where
$|G_{\rm{m}}(k)| - G_{\rm gr}(k)$ becomes maximal.
\begin{figure}[ht]
    \includegraphics[width=85mm]{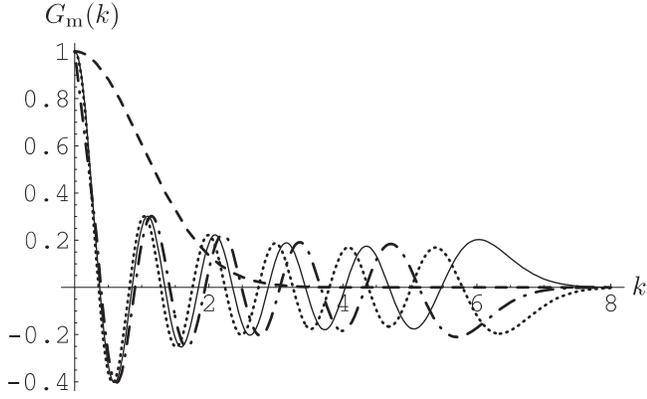}
\caption{Characteristic functions $G_{\rm{m}}(k)$  versus $k$ for 
number states $m=10$ (full line), $9$
  (dashed-dotted) and $11$ (dotted), together with the
  classical limit $G_{\rm gr}(k)=e^{-k^2/2}$ (dashed).}
\end{figure}

Now we consider the time evolution of the characteristic function,
$G_{\rm{m}}(k,t)$, caused by the thermal reservoir.  The initial preparation
of a Fock state $|m\rangle$ allows one to distinguish motional-state
redistribution effects from dephasing effects.  Using Eqs~(\ref{char-eq}) to
(\ref{CF-n}), we obtain 
\begin{eqnarray}\label{Phi-k(t)}
G_{\rm{m}}(k,t)& =  & 
\exp\left\{-\bar{n}[1-\exp(-2\gamma t)]k^2\right\}
 \nonumber \\
& & \times L_m \left(k^2\,e^{-2\gamma t} \right)\,\exp{(-k^2/2)} .
\end{eqnarray}
This function can be monitored for a chosen time $t$, as discussed in
connection with Eq.~(\ref{eq:ini}).

To get more insight into the time evolution, we consider
the time derivative of the characteristic function at $t=0$,
\begin{equation}
 \dot{G}_{\rm{m}}(k,0)= 2\gamma
[L_{m-1}^{(1)}(k^2)-\bar{n} L_m(k^2)]\,k^2\,e^{-k^2/2}\,.
\end{equation}
It strongly depends on both $k$ and $\bar{n}$, for $m=10$ the maximum of the
time derivative occurs for $k=k_{\rm max} \approx 6$. Hence this argument of
the characteristic function will be of particular interest to detect the
decoherence in a sensitive way. Comparing with Fig.~1, the most sensitive
reaction on the motional-state redistributions occurs around the outermost
maximum of the initial characteristic function, where the nonclassical
features are dominant. This reflects the expected high fragility of
nonclassical effects.

For the used Fock state it is interesting to compare the time evolution of the
characteristic function with that of the
sub-Poissonian statistics. For the
Mandel parameter, $Q=(\langle(\Delta\hat{n})^2\rangle
-\langle\hat{n}\rangle)/\langle\hat{n}\rangle$, which measures the deviation
from the Poissonian statistics, we get
\begin{equation}\label{Q}
Q(t)=\frac{(\bar{n}^2
-2\bar{n}m-m)e^{-4\gamma t}
  +2\bar{n}(m-\bar{n})e^{-2\gamma t} +\bar{n}^2}{me^{-2\gamma t}
  +\bar{n}(1-e^{-2\gamma t})}\, .
\end{equation}
Negative values of $Q$ characterize a nonclassical quantum state showing
sub-Poissonian statistics. 
\begin{figure}[h!]
    \includegraphics[width=85mm]{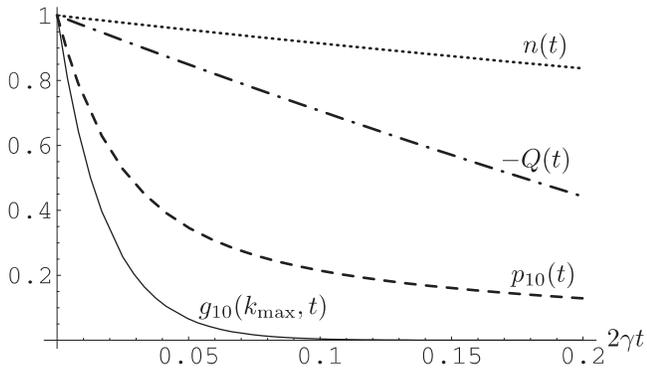}
\caption{The time evolution is shown for the occupation probability
  $p_{10}(t)$ of the state $|m=10 \rangle$, the
   normalized mean excitation number $n(t)$, the Mandel parameter $Q(t)$, and
  the normalized characteristic function $g_{10}(k_{\rm max},t)$, for
  $\bar{n}=1$.}
\end{figure}

In Fig.~2 we show the time evolution of the normalized characteristic
function, $g_{\rm{10}}(k,t)=G_{\rm{10}}(k,t)/G_{\rm{10}}(k,0)$, for $k=k_{\rm
  max}$ and $\bar{n}=1$.  For comparison, the time evolutions are also shown
for the Mandel $Q$ parameter, the occupation probability $p_{10}(t)$ of the
initially prepared Fock state $|m=10\rangle$, and the normalized mean
motional-state excitation $n(t)$, $n(t)=
\langle\hat{n}(t)\rangle/\langle\hat{n}(0)\rangle$ with
$\langle\hat{n}(t)\rangle = m e^{-2\gamma t}+\bar{n}(1-e^{2\gamma t})$. It is
clearly seen that the nonclassical characteristic function shows the fastest
decay and evolves much faster than the nonclassical property described by the
Mandel $Q$ parameter.  In fact, the nonclassical part of the characteristic
function decays even faster than the occupation probability of the initially
prepared Fock state. Thus it yields a highly sensitive means of detection,
since even tiny dissipation leads to a noticeable effect on that part of the
characteristic function.

How can we explain the highly sensitive behavior of the
characteristic function in its outermost maximum? 
The quadrature characteristic function, 
\begin{equation}\label{G-sum}
G_{\rm 10}(k,t)=\sum_{n=0}^{\infty}
p_{ n}(t)\, G_{\rm n}(k),
\end{equation}
can be written as a sum of characteristic functions of number states weighted
with the occupation probabilities $p_{n}(t)$.  From the master
equation~(\ref{master-eq}) it follows that at the very beginning of the time
evolution the only nonvanishing occupation probabilities are $p_9(t)$,
$p_{11}(t)$ and $p_{10}(t)= 1- p_9(t)-p_{11}(t)$.  It is seen from Fig.~1 that
$G_{\rm{}9}(k)$ and $G_{\rm{11}}(k)$ attain for $k=k_{\rm{max}}$ roughly the
same absolute value as $G_{\rm{10}}(k)$, but with opposite sign.  Thus,
according to Eq.~(\ref{G-sum}), the increase of $p_9(t)$ and $p_{11}(t)$ leads
to a faster decay of $G_{\rm{10}}(k,t)$ compared with $p_{10}(t)$.  The decay of
$G_{\rm 10}(k_{\rm max},t)$ is roughly twice as fast as the decay of
$p_{10}(t)$.  Hence the detection of the characteristic function is both
simpler and more sensitive than a number-state measurement, even though the
latter is directly related to the prepared number state.

For comparison we may look at the characteristic function for other $k$-values, which are
within the classical region (such as $k\approx 1)$ or which violate the
classical limit only slightly (e.g.  $k\approx 2$). Then the temporal
evolution is significantly slower. For $k \approx 1$ the neighboring functions
are almost equal: $G_9(k) \approx G_{10}(k) \approx G_{11}(k)$, cf. Fig.~1.
Thus increasing values of $p_9$ and $p_{11}$ nearly compensate the decay of
$p_{10}$, resulting in a slow decay of $G_{10}(k,t)$.

\subsection{Decoherence of Radiation Fields}

In the following we will consider a typical application of our method that
could be realized for a radiation field.  The aim is to identify small
decoherence effects during the transmission of nonclassical light through
media. This is of great importance for applications such as in quantum
communication, where the transmission can be performed via optical fibers or
through the atmosphere.

The preparation of photon number states, at least for larger photon numbers,
is much more difficult to realize as the Fock-state preparation in the ion
trap. However, we will see that squeezed light can be used as the nonclassical
radiation source for our measurement principle as well. Let us consider a
squeezed vacuum,
\begin{equation}
|{\rm sv}\rangle
=\exp[-(r/2)\hat{a}^{\dagger 2} +(r/2)\hat{a}^2]|0\rangle,
\end{equation}
with $r\geq 0$.  The action of the medium is modeled by the thermal
reservoir as before.

Due to the contact  with the reservoir, the minimum of the quadrature
variance ($\varphi=0$ in Eq.~(\ref{eq:xop})) behaves like 
\begin{equation}
\langle (\Delta \hat{x}(t))^2 \rangle_{\rm min} = (1+2\bar{n})(1-e^{-2\gamma t}) +
e^{-2r}e^{-2\gamma t},
\end{equation}
where $t$ represents the propagation time through the
medium.  For the chosen phase, $\varphi=0$, the nonclassical effect is most
pronounced. The characteristic function for this phase is simply given by
\begin{equation}
G_{\rm{sv}}(k,t)=  e^{-k^2\langle (\Delta \hat{x}(t))^2 \rangle_{\rm min}/2}.
\end{equation}
Its time derivative yields the $k$-value with the
maximum sensitivity of our method to be 
\begin{equation}
k_{\rm max} =\sqrt{2e^{2r}}\,,
\end{equation} 
which is independent of the value of $\bar{n}$.
\begin{figure}[t]
 \includegraphics[width=85mm]{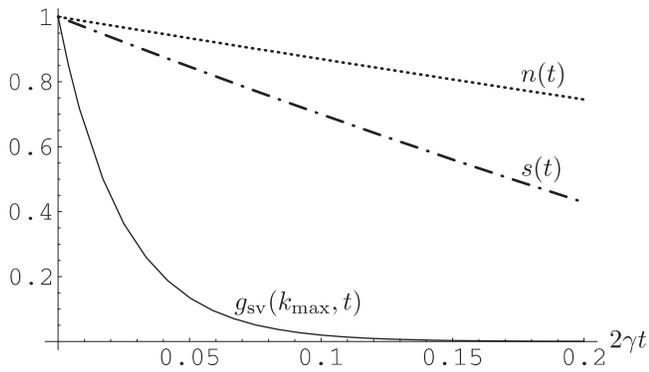}
\caption{Time evolution of the normalized mean photon number $n(t)$, the
  quadrature variance $s(t)$, and the characteristic function
  $g_{\rm{sv}}(k_{\rm max},t)$ of an initial squeezed vacuum state with
  $r=1.32$, for $\bar{n}=1$.}
\end{figure}

In Fig.~3 we show the time evolution of the normalized mean photon number
$n(t)$, the normally-ordered quadrature variance $s(t)$, and the
characteristic function $g_{\rm{sv}}(k,t)$, where $s(t)\! = \!\langle
:\!(\Delta \hat{x}(t))^2\!: \rangle_{\rm min}/ \langle :\!(\Delta
\hat{x}(0))^2\!: \rangle_{\rm min}$ and
$g_{\rm{sv}}(k,t)=G_{\rm{sv}}(k,t)/G_{\rm{sv}}(k,0)$.  As expected, the
fastest decay is observed for the characteristic function at $k=k_{\rm max}$.
To detect this highly sensitive reaction on the reservoir effects, the
characteristic function can be sampled, at the end of the transmission
channel, via balanced homodyne detection. For the sensitive detection method
under study, we only need to consider the behavior for $k=k_{\rm max}$ and for
the phase with the minimal quadrature variance.

\section{Summary}

In conclusion, we have shown that highly sensitive measurements can be
performed by detecting the strongly nonclassical part of the quadrature
characteristic function. The method makes use of the high fragility of the
nonclassical effects. For example, the direct observation of the
characteristic function can monitor the redistribution of the motional-state
occupations of an initially prepared Fock state of a trapped ion. Sensitive
optical decoherence measurements can be realized by using squeezed light.  The
method may be useful for highly sensitive noise control in quantum information
systems. It is based on a universal measurement principle, that can be used  
with different types of initially prepared nonclassical states.

\end{document}